\documentclass[preprint,preprintnumbers,amsmath,amssymb,superscriptaddress]{revtex4}

\usepackage{graphicx}
\usepackage{dcolumn}
\usepackage{bm}

\makeatletter
\def\@seccntformat#1{}
\makeatother
\renewcommand{\numberline}[1]{}

\begin{document}

\centerline {NATURE \textbf{452}, 970 (2008) [News \& Views at
NAT.PHYS. 4, 348 (2008)]}

\title{A topological Dirac insulator in a quantum spin Hall phase}

\author{D. Hsieh}
\affiliation{Joseph Henry Laboratories of Physics, Department of Physics, Princeton
University, Princeton, NJ 08544, USA}

\author{D. Qian}
\affiliation{Joseph Henry Laboratories of Physics, Department of Physics, Princeton
University, Princeton, NJ 08544, USA}

\author{L. Wray}
\affiliation{Joseph Henry Laboratories of Physics, Department of Physics, Princeton
University, Princeton, NJ 08544, USA}

\author{Y. Xia}
\affiliation{Joseph Henry Laboratories of Physics, Department of Physics, Princeton
University, Princeton, NJ 08544, USA}
\affiliation{Princeton Center
for Complex Materials, Princeton University, Princeton NJ 08544,
USA}

\author{Y. S. Hor}
\affiliation{Department of Chemistry, Princeton University,
Princeton, NJ 08544, USA}

\author{R. J. Cava}
\affiliation{Department of Chemistry, Princeton University,
Princeton, NJ 08544, USA}

\author{M. Z. Hasan}
\affiliation{Joseph Henry Laboratories of Physics, Department of Physics, Princeton
University, Princeton, NJ 08544, USA}
\affiliation{Princeton Center
for Complex Materials, Princeton University, Princeton NJ 08544,
USA}
\affiliation{Princeton Institute for the Science and Technology of Materials (PRISM), Princeton University, Princeton NJ 08544,
USA}
\email{mzhasan@Princeton.edu}




\maketitle

\textbf{When electrons are subject to a large external magnetic
field, the conventional charge quantum Hall effect
\cite{Klitzing,Tsui} dictates that an electronic excitation gap is
generated in the sample bulk, but metallic conduction is permitted
at the boundary. Recent theoretical models suggest that certain bulk
insulators with large spin-orbit interactions may also naturally
support conducting topological boundary states in the extreme quantum limit
\cite{Kane(Graphene),Bernevig(QSHE),Sheng(QSHE)}, which opens up the
possibility for studying unusual quantum Hall-like phenomena in zero
external magnetic fields \cite{Haldane(P-anomaly)}. Bulk
Bi$_{1-x}$Sb$_x$ single crystals are predicted to be prime
candidates \cite{Fu:STI1,Murukami} for one such unusual Hall phase
of matter known as the topological insulator \cite{Fu:STI2,
Moore:STI1, Roy}. The hallmark of a topological insulator is the
existence of metallic surface states that are higher dimensional
analogues of the edge states that characterize a quantum spin Hall
insulator
\cite{Kane(Graphene),Bernevig(QSHE),Sheng(QSHE),Haldane(P-anomaly),Fu:STI1,
Murukami, Fu:STI2, Moore:STI1, Roy, Bernevig:HgTe, Konig}. In
addition to its interesting boundary states, the bulk of
Bi$_{1-x}$Sb$_x$ is predicted to exhibit three-dimensional Dirac
particles \cite{Wolff, Fukuyama, Buot, Lenoir}, another topic of
heightened current interest following the new findings of
two-dimensional graphene \cite{Zhang, Novoselov, Zhou} and charge
quantum Hall fractionalization observed in pure bismuth
\cite{Behnia}. However, despite numerous transport and magnetic
measurements on the Bi$_{1-x}$Sb$_x$ family since the 1960s
\cite{Lenoir}, no direct evidence of either topological quantum Hall-like states
or bulk Dirac particles has ever been found. Here, using
incident-photon-energy-modulated angle-resolved photoemission
spectroscopy (IPEM-ARPES), we report the direct observation of
massive Dirac particles in the bulk of Bi$_{0.9}$Sb$_{0.1}$, locate
the Kramers' points at the sample's boundary and provide a
comprehensive mapping of the topological Dirac insulator's gapless surface
modes. These findings taken together suggest that the observed surface state on the boundary of the bulk insulator is a realization of the much sought exotic ``topological metal'' \cite{Fu:STI2, Moore:STI1, Roy}. They also suggest that this material has potential application in developing next-generation quantum computing devices that may
incorporate ``light-like" bulk carriers and topologically protected spin-textured edge-surface currents.}

\newpage

Bismuth is a semimetal with strong spin-orbit interactions. Its band
structure is believed to feature an indirect negative gap between
the valence band maximum at the T point of the bulk Brillouin zone
(BZ) and the conduction band minima at three equivalent L points
\cite{Lenoir,Liu} (here we generally refer to these as a single
point, L). The valence and conduction bands at L are derived from
antisymmetric (L$_a$) and symmetric (L$_s$) $p$-type orbitals,
respectively, and the effective low-energy Hamiltonian at this point
is described by the (3+1)-dimensional relativistic Dirac equation
\cite{Wolff, Fukuyama, Buot}. The resulting dispersion relation,
$E(\vec{k}) = \pm \sqrt{ {(\vec{v} \cdot \vec{k})}^2 + \Delta^2}
\approx \vec{v} \cdot \vec{k}$, is highly linear owing to the
combination of an unusually large band velocity $\vec{v}$ and a
small gap $\Delta$ (such that $\lvert \Delta / \lvert \vec{v} \rvert
\rvert \approx 5 \times 10^{-3} $\AA$^{-1}$) and has been used to
explain various peculiar properties of bismuth \cite{Wolff,
Fukuyama, Buot}. Substituting bismuth with antimony is believed to
change the critical energies of the band structure as follows (see
Fig.1). At an Sb concentration of $x \approx 4\%$, the gap $\Delta$
between L$_a$ and L$_s$ closes and a massless three-dimensional (3D)
Dirac point is realized. As $x$ is further increased this gap
re-opens with inverted symmetry ordering, which leads to a change in
sign of $\Delta$ at each of the three equivalent L points in the BZ.
For concentrations greater than $x \approx 7\%$ there is no overlap
between the valence band at T and the conduction band at L, and the
material becomes an inverted-band insulator. Once the band at T
drops below the valence band at L, at $x \approx 8\%$, the system
evolves into a direct-gap insulator whose low energy physics is
dominated by the spin-orbit coupled Dirac particles at L
\cite{Fu:STI1, Lenoir}.

Recently, semiconductors with inverted band gaps have been proposed
to manifest the two-dimensional (2D) quantum spin Hall phase, which
is predicted to be characterized by the presence of metallic 1D edge
states
\cite{Kane(Graphene),Bernevig(QSHE),Sheng(QSHE),Bernevig:HgTe}.
Although a band-inversion mechanism and edge states have been
invoked to interpret the transport results in 2D mercury telluride
semiconductor quantum wells \cite{Konig}, no 1D edge states are
directly imaged, so their topological character is unknown. Recent
theoretical treatments \cite{Fu:STI1, Murukami} have focused on the
strongly spin-orbit coupled, band-inverted Bi$_{1-x}$Sb$_x$ series
as a possible 3D bulk realization of the quantum spin Hall phase in
which the 1D edge states are expected to take the form of 2D surface
states \cite{Fu:STI1, Murukami, Fu:STI2} that may be directly imaged
and spectroscopically studied, making it feasible to identify their
topological order parameter character. Most importantly, the 3D
phase is a new phase of matter in terms of its topological
distinctions \cite{Moore:STI1}.

High-momentum-resolution angle-resolved photoemission spectroscopy
performed with varying incident photon energy (IPEM-ARPES) allows
for measurement of electronic band dispersion along various momentum
space ($\vec{k}$-space) trajectories in the 3D bulk BZ. ARPES
spectra taken along two orthogonal cuts through the L point of the
bulk BZ of Bi$_{0.9}$Sb$_{0.1}$ are shown in Figs 1a and c. A
$\Lambda$-shaped dispersion whose tip lies less than 50 meV below
the Fermi energy ($E_F$) can be seen along both directions.
Additional features originating from surface states that do not
disperse with incident photon energy are also seen. Owing to the
finite intensity between the bulk and surface states, the exact
binding energy ($E_B$) where the tip of the $\Lambda$-shaped band
dispersion lies is unresolved. The linearity of the bulk
$\Lambda$-shaped bands is observed by locating the peak positions at
higher $E_B$ in the momentum distribution curves (MDCs), and the
energy at which these peaks merge is obtained by extrapolating
linear fits to the MDCs. Therefore 50 meV represents a lower bound
on the energy gap $\Delta$ between L$_a$ and L$_s$. The magnitude of
the extracted band velocities along the $k_x$ and $k_y$ directions
are $7.9 \pm 0.5 \times 10^4$ ms$^{-1}$ and $10.0 \pm 0.5 \times
10^5$ ms$^{-1}$, respectively, which are similar to the tight
binding values $7.6 \times 10^4$ ms$^{-1}$ and $9.1 \times 10^5$
ms$^{-1}$ calculated for the L$_a$ band of bismuth \cite{Liu}. Our
data are consistent with the extremely small effective mass of
$0.002m_e$ (where $m_e$ is the electron mass) observed in
magneto-reflection measurements on samples with $x = 11\%$
\cite{Hebel}. The Dirac point in graphene, co-incidentally, has a
band velocity ($|v_F| \approx 10^6$ ms$^{-1}$) \cite{Zhang}
comparable to what we observe for Bi$_{0.9}$Sb$_{0.1}$, but its
spin-orbit coupling is several orders of magnitude weaker
\cite{Kane(Graphene)}, and the only known method of inducing a gap
in the Dirac spectrum of graphene is by coupling to an external
chemical substrate \cite{Zhou}. The Bi$_{1-x}$Sb$_x$ series thus
provides a rare opportunity to study relativistic Dirac Hamiltonian
physics in a 3D condensed matter system where the intrinsic (rest)
mass gap can be easily tuned.

Studying the band dispersion perpendicular to the sample surface
provides a way to differentiate bulk states from surface states in a
3D material. To visualize the near-$E_F$ dispersion along the 3D L-X
cut (X is a point that is displaced from L by a $k_z$ distance of
3$\pi/c$, where $c$ is the lattice constant), in Fig.2a we plot
energy distribution curves (EDCs), taken such that electrons at
$E_F$ have fixed in-plane momentum $(k_x, k_y)$ = (L$_x$, L$_y$) =
(0.8 \AA$^{-1}$, 0.0 \AA$^{-1}$), as a function of photon energy
($h\nu$). There are three prominent features in the EDCs: a
non-dispersing, $k_z$ independent, peak centered just below $E_F$ at
about $-$0.02 eV; a broad non-dispersing hump centered near $-$0.3
eV; and a strongly dispersing hump that coincides with the latter
near $h\nu$ = 29 eV. To understand which bands these features
originate from, we show ARPES intensity maps along an in-plane cut
$\bar{K} \bar{M} \bar{K}$ (parallel to the $k_y$ direction) taken
using $h\nu$ values of 22 eV, 29 eV and 35 eV, which correspond to
approximate $k_z$ values of L$_z -$ 0.3 \AA$^{-1}$, L$_z$, and L$_z$
+ 0.3 \AA$^{-1}$ respectively (Fig.2b). At $h\nu$ = 29 eV, the low
energy ARPES spectral weight reveals a clear $\Lambda$-shaped band
close to $E_F$. As the photon energy is either increased or
decreased from 29 eV, this intensity shifts to higher binding
energies as the spectral weight evolves from the $\Lambda$-shaped
into a $\cup$-shaped band. Therefore the dispersive peak in Fig.2a
comes from the bulk valence band, and for $h\nu$ = 29 eV the high
symmetry point L = (0.8, 0, 2.9) appears in the third bulk BZ. In
the maps of Fig.2b with respective $h\nu$ values of 22 eV and 35 eV,
overall weak features near $E_F$ that vary in intensity remain even
as the bulk valence band moves far below $E_F$. The survival of
these weak features over a large photon energy range (17 to 55 eV)
supports their surface origin. The non-dispersing feature centered
near $-0.3$ eV in Fig.2a comes from the higher binding energy
(valence band) part of the full spectrum of surface states, and the
weak non-dispersing peak at $-0.02$ eV reflects the low energy part
of the surface states that cross $E_F$ away from the $\bar{M}$ point
and forms the surface Fermi surface (Fig.2c).

Having established the existence of an energy gap in the bulk state
of Bi$_{0.9}$Sb$_{0.1}$ (Figs 1 and 2) and observed linearly
dispersive bulk bands uniquely consistent with strong spin-orbit
coupling model calculations \cite{Wolff, Fukuyama, Buot, Liu} (see
Supplementary Information for full comparison with theoretical
calculation), we now discuss the topological character of its
surface states, which are found to be gapless (Fig.2c). In general,
the states at the surface of spin-orbit coupled compounds are
allowed to be spin split owing to the loss of space inversion
symmetry $[E(k,\uparrow) = E(-k,\uparrow)]$. However, as required by
Kramers' theorem, this splitting must go to zero at the four time
reversal invariant momenta (TRIM) in the 2D surface BZ. As discussed
in \cite{Fu:STI1, Fu:STI2}, along a path connecting two TRIM in the
same BZ, the Fermi energy inside the bulk gap will intersect these
singly degenerate surface states either an even or odd number of
times. When there are an even number of surface state crossings, the
surface states are topologically trivial because weak disorder (as
may arise through alloying) or correlations can remove \emph{pairs}
of such crossings by pushing the surface bands entirely above or
below $E_F$. When there are an odd number of crossings, however, at
least one surface state must remain gapless, which makes it
non-trivial \cite{Fu:STI1, Murukami, Fu:STI2}. The existence of such
topologically non-trivial surface states can be theoretically
predicted on the basis of the \emph{bulk} band structure only, using
the $Z_2$ invariant that is related to the quantum Hall Chern number
\cite{Kane(QSHE-Z2)}. Materials with band structures with $Z_2 = +1$
are ordinary Bloch band insulators that are topologically equivalent
to the filled shell atomic insulator, and are predicted to exhibit
an even number (including zero) of surface state crossings.
Materials with bulk band structures with $Z_2 = -1$ on the other
hand, which are expected to exist in rare systems with strong
spin-orbit coupling acting as an internal quantizing magnetic field
on the electron system \cite{Haldane(P-anomaly)}, and inverted bands
at an odd number of high symmetry points in their bulk 3D BZs, are
predicted to exhibit an odd number of surface state crossings,
precluding their adiabatic continuation to the atomic insulator
\cite{Kane(Graphene), Fu:STI1, Murukami, Fu:STI2, Moore:STI1, Roy,
Bernevig:HgTe, Konig}. Such ``topological quantum Hall metals''
\cite{Fu:STI2, Moore:STI1, Roy} cannot be realized in a purely 2D
electron gas system such as the one realized at the interface of
GaAs/GaAlAs systems.

In our experimental case, namely the (111) surface of
Bi$_{0.9}$Sb$_{0.1}$, the four TRIM are located at $\bar{\Gamma}$
and three $\bar{M}$ points that are rotated by $60^{\circ}$ relative
to one another. Owing to the three-fold crystal symmetry (A7 bulk
structure) and the observed mirror symmetry of the surface Fermi
surface across $k_x = 0$ (Fig.2), these three $\bar{M}$ points are
equivalent (and we henceforth refer to them as a single point,
$\bar{M}$). The mirror symmetry $[E(k_y) = E(-k_y)]$ is also
expected from time reversal invariance exhibited by the system. The
complete details of the surface state dispersion observed in our
experiments along a path connecting $\bar{\Gamma}$ and $\bar{M}$ are
shown in Fig.3a; finding this information is made possible by our
experimental separation of surface states from bulk states. As for
bismuth (Bi), two surface bands emerge from the bulk band continuum
near $\bar{\Gamma}$ to form a central electron pocket and an
adjacent hole lobe \cite{Ast:Bi1, Hochst,Hofmann}. It has been
established that these two bands result from the spin-splitting of a
surface state and are thus singly degenerate \cite{Hirahara,
Hofmann}. On the other hand, the surface band that crosses $E_F$ at
$-k_x \approx 0.5$ \AA$^{-1}$, and forms the narrow electron pocket
around $\bar{M}$, is clearly doubly degenerate, as far as we can
determine within our experimental resolution. This is indicated by
its splitting below $E_F$ between $-k_x \approx 0.55$ \AA$^{-1}$ and
$\bar{M}$, as well as the fact that this splitting goes to zero at
$\bar{M}$ in accordance with Kramers theorem. In semimetallic single
crystal bismuth, only a single surface band is observed to form the
electron pocket around $\bar{M}$ \cite{Hengsberger, Ast:Bi2}.
Moreover, this surface state overlaps, hence becomes degenerate
with, the bulk conduction band at L (L projects to the surface
$\bar{M}$ point) owing to the semimetallic character of Bi (Fig.3b).
In Bi$_{0.9}$Sb$_{0.1}$ on the other hand, the states near $\bar{M}$
fall completely inside the bulk energy gap preserving their purely
surface character at $\bar{M}$ (Fig.3a). The surface Kramers doublet
point can thus be defined in the bulk insulator (unlike in Bi
\cite{Hirahara,Ast:Bi1, Hochst, Hofmann, Hengsberger, Ast:Bi2}) and
is experimentally located in Bi$_{0.9}$Sb$_{0.1}$ samples to lie
approximately 15 $\pm$ 5 meV below $E_F$ at $\vec{k} = \bar{M}$
(Fig.3a). For the precise location of this Kramers point, it is
important to demonstrate that our alignment is strictly along the
$\bar{\Gamma} - \bar{M}$ line. To do so, we contrast high resolution
ARPES measurements taken along the $\bar{\Gamma} - \bar{M}$ line
with those that are slightly offset from it (Fig.3e). Figs 3f-i show
that with $k_y$ offset from the Kramers point at $\bar{M}$ by less
than 0.02 \AA$^{-1}$, the degeneracy is lifted and only one band
crosses $E_F$ to form part of the bow-shaped electron distribution
(Fig.3d). Our finding of five surface state crossings (an odd rather
than an even number) between $\bar{\Gamma}$ and $\bar{M}$ (Fig.3a),
confirmed by our observation of the Kramers degenerate point at the
TRIM, indicates that these gapless surface states are topologically
non-trivial. This corroborates our bulk electronic structure result
that Bi$_{0.9}$Sb$_{0.1}$ is in the insulating band-inverted ($Z_2 =
-1$) regime (Fig.3c), which contains an odd number of bulk (gapped)
Dirac points, and is topologically analogous to an integer quantum
spin Hall insulator.

Our experimental results taken collectively strongly suggest that
Bi$_{0.9}$Sb$_{0.1}$ is quite distinct from graphene \cite{Zhang,
Novoselov} and represents a novel state of quantum matter: a
strongly spin-orbit coupled insulator with an odd number of Dirac
points with a negative $Z_2$ topological Hall phase, which realizes
the ``parity anomaly without Fermion doubling". Our work further
demonstrates a general methodology for possible future
investigations of \emph{novel topological orders} in exotic quantum
matter.

\textbf{Note Added :
In a very recent work we have successfully imaged the spin-polarization of the topological edge modes using high-resolution spin-resolved-ARPES \cite{Science}.}

\vspace{1cm}

\small{\textbf{Acknowledgements} We thank P. W. Anderson, B. A.
Bernevig, F. D. M. Haldane, D. A. Huse, C. L. Kane, R. B. Laughlin,
N. P. Ong, A. N. Pasupathy and D. C. Tsui for discussions. This work is supported by the DOE Office of Basic Energy Science and materials synthesis is supported by the NSF MRSEC.}

\small{\textbf{Author information} Correspondence and requests for
materials should be addressed to M.Z.H (mzhasan@princeton.edu).}

\section{Methods}

High resolution IPEM-ARPES data have been taken at Beamlines 12.0.1
and 10.0.1 of the Advanced Light Source in Lawrence Berkeley
National Laboratory, as well as at PGM Beamline of the Synchrotron
Radiation Center in Wisconsin, with photon energies from 17 to 55 eV
and energy resolution from 9 to 40 meV and momentum (k-)resolution
better than $1.5 \%$ of the surface Brillouin zone. Data were taken
on high quality bulk single crystal Bi$_{1-x}$Sb$_x$ at a
temperature of 15 K and chamber pressures better than $8 \times
10^{-11}$ torr. Throughout this paper, the bulk bands presented are
from those measured in the third bulk Brillouin zone to ensure a
high degree of signal-to-noise contrast, and the $k_z$ values are
estimated using the standard free-electron final state
approximation.


\newpage

\textbf{Figure 1. Dirac-like dispersion near the L point in the bulk
Brillouin zone.} Selected ARPES intensity maps of
Bi$_{0.9}$Sb$_{0.1}$ are shown along three $\vec{k}$-space cuts
through the L point of the bulk 3D Brillouin zone (BZ). The
presented data are taken in the third BZ with L$_z$ = 2.9 \AA$^{-1}$
with a photon energy of 29 eV. The cuts are along \textbf{a}, the
$k_y$ direction, \textbf{b}, a direction rotated by approximately
$10^{\circ}$ from the $k_y$ direction, and \textbf{c}, the $k_x$
direction. Each cut shows a $\Lambda$-shaped bulk band whose tip
lies below the Fermi level signalling a bulk gap. The surface states
are denoted SS and are all identified in Fig.2 (for further
identification via theoretical calculations see Supplementary
Information). \textbf{d}, Momentum distribution curves (MDCs)
corresponding to the intensity map in \textbf{a}. \textbf{f}, Log
scale plot of the MDCs corresponding to the intensity map in
\textbf{c}. The red lines are guides to the eye for the bulk
features in the MDCs. \textbf{e}, Schematic of the bulk 3D BZ of
Bi$_{1-x}$Sb$_x$ and the 2D BZ of the projected (111) surface. The
high symmetry points $\bar{\Gamma}$, $\bar{M}$ and $\bar{K}$ of the
surface BZ are labeled. Schematic evolution of bulk band energies as
a function of $x$ is shown. The L band inversion transition occurs
at $x \approx 0.04$, where a 3D gapless Dirac point is realized, and
the composition we study here (for which $x = 0.1$) is indicated by
the green arrow. A more detailed phase diagram based on our
experiments is shown in Fig.3c.

\newpage

\textbf{Figure 2. Dispersion along the cut in the
$\mathbf{k_z}$-direction.} Surface states are experimentally
identified by studying their out-of-plane momentum dispersion
through the systematic variation of incident photon energy.
\textbf{a}, Energy distribution curves (EDCs) of
Bi$_{0.9}$Sb$_{0.1}$ with electrons at the Fermi level ($E_F$)
maintained at a fixed in-plane momentum of ($k_x$=0.8 \AA$^{-1}$,
$k_y$=0.0 \AA$^{-1}$) are obtained as a function of incident photon
energy to identify states that exhibit no dispersion perpendicular
to the (111)-plane along the direction shown by the double-headed
arrow labeled ``3" in the inset (see Methods for detailed
procedure). Selected EDC data sets with photon energies of 28 eV to
32 eV in steps of 0.5 eV are shown for clarity. The non-energy
dispersive ($k_z$ independent) peaks near $E_F$ are the surface
states (SS). \textbf{b}, ARPES intensity maps along cuts parallel to
$k_y$ taken with electrons at $E_F$ fixed at $k_x$ = 0.8 \AA$^{-1}$
with respective photon energies of $h \nu$ = 22 eV, 29 eV and 35 eV
(for a conversion map from photon energy to $k_z$ see Supplementary
Information). The faint $\Lambda$-shaped band at $h \nu$ = 22 eV and
$h \nu$ = 35 eV shows some overlap with the bulk valence band at L
($h \nu$ = 29 eV), suggesting that it is a resonant surface state
degenerate with the bulk state in some limited k-range near $E_F$.
The flat band of intensity centered about $-$2 eV in the $h \nu$ =
22 eV scan originates from Bi 5d core level emission from second
order light. \textbf{c}, Projection of the bulk BZ (black lines)
onto the (111) surface BZ (green lines). Overlay (enlarged in inset)
shows the high resolution Fermi surface (FS) of the metallic SS
mode, which was obtained by integrating the ARPES intensity (taken
with $h \nu$ = 20 eV) from $-$15 meV to 10 meV relative to $E_F$.
The six tear-drop shaped lobes of the surface FS close to
$\bar{\Gamma}$ (center of BZ) show some intensity variation between
them that is due to the relative orientation between the axes of the
lobes and the axis of the detector slit. The six-fold symmetry was
however confirmed by rotating the sample in the $k_x-k_y$ plane.
EDCs corresponding to the cuts A, B and C are also shown; these
confirm the gapless character of the surface states in bulk
insulating Bi$_{0.9}$Sb$_{0.1}$.

\newpage

\textbf{Figure 3. The topological gapless surface states in bulk
insulating Bi$_{0.9}$Sb$_{0.1}$.} \textbf{a}, The surface band
dispersion second derivative image (SDI) of Bi$_{0.9}$Sb$_{0.1}$
along $\bar{\Gamma} - \bar{M}$. The shaded white area shows the
projection of the bulk bands based on ARPES data, as well as a rigid
shift of the tight binding bands to sketch the unoccupied bands
above the Fermi level. To maintain high momentum resolution, data were collected in two segments of momentum space, then the intensities were normalized using background level above the Fermi level. A non-intrinsic flat band of intensity near
$E_F$ generated by the SDI analysis was rejected to isolate the
intrinsic dispersion. The Fermi crossings of the surface state are
denoted by yellow circles, with the band near $-k_x \approx 0.5$
\AA$^{-1}$ counted twice owing to double degeneracy. The red lines
are guides to the eye. An in-plane rotation of the sample by
$60^{\circ}$ produced the same surface state dispersion. The EDCs
along $\bar{\Gamma} - \bar{M}$ are shown to the right. There are a
total of five crossings from $\bar{\Gamma} - \bar{M}$ which
indicates that these surface states are topologically non-trivial.
The number of surface state crossings in a material (with an odd
number of Dirac points) is related to the topological $Z_2$
invariant (see text). \textbf{b}, The resistivity curves of Bi and
Bi$_{0.9}$Sb$_{0.1}$ reflect the contrasting transport behaviours.
The presented resistivity curve for pure bismuth has been multiplied
by a factor of 80 for clarity. \textbf{c}, Schematic variation of
bulk band energies of Bi$_{1-x}$Sb$_x$ as a function of $x$ (based
on band calculations and on \cite{Fu:STI1, Lenoir}).
Bi$_{0.9}$Sb$_{0.1}$ is a direct gap bulk Dirac point insulator well
inside the inverted-band regime, and its surface forms a
``topological metal'' - the 2D analogue of the 1D edge states in
quantum spin Hall systems. \textbf{d}, ARPES intensity integrated
within $\pm 10$ meV of $E_F$ originating solely from the surface
state crossings. The image was plotted by stacking along the
negative $k_x$ direction a series of scans taken parallel to the
$k_y$ direction. \textbf{e}, Outline of Bi$_{0.9}$Sb$_{0.1}$ surface
state ARPES intensity near $E_F$ measured in \textbf{d}. White lines
show scan directions ``1'' and ``2''. \textbf{f}, Surface band
dispersion along direction ``1'' taken with $h \nu$ = 28 eV and the
corresponding EDCs (\textbf{g}). The surface Kramers degenerate
point, critical in determining the topological $Z_2$ class of a band
insulator, is clearly seen at $\bar{M}$, approximately $15 \pm 5$
meV below $E_F$. (We note that the scans are taken along the
negative $k_x$ direction, away from the bulk L point.) \textbf{h},
Surface band dispersion along direction ``2'' taken with $h \nu$
 = 28 eV and the corresponding EDCs (\textbf{i}). This scan no longer
passes through the $\bar{M}$-point, and the observation of two well
separated bands indicates the absence of Kramers degeneracy as
expected, which cross-checks the result in (\textbf{a}).

\newpage

\begin{figure}
\includegraphics[scale=0.75,clip=true, viewport=0.0in 0in 8.5in 7.7in]{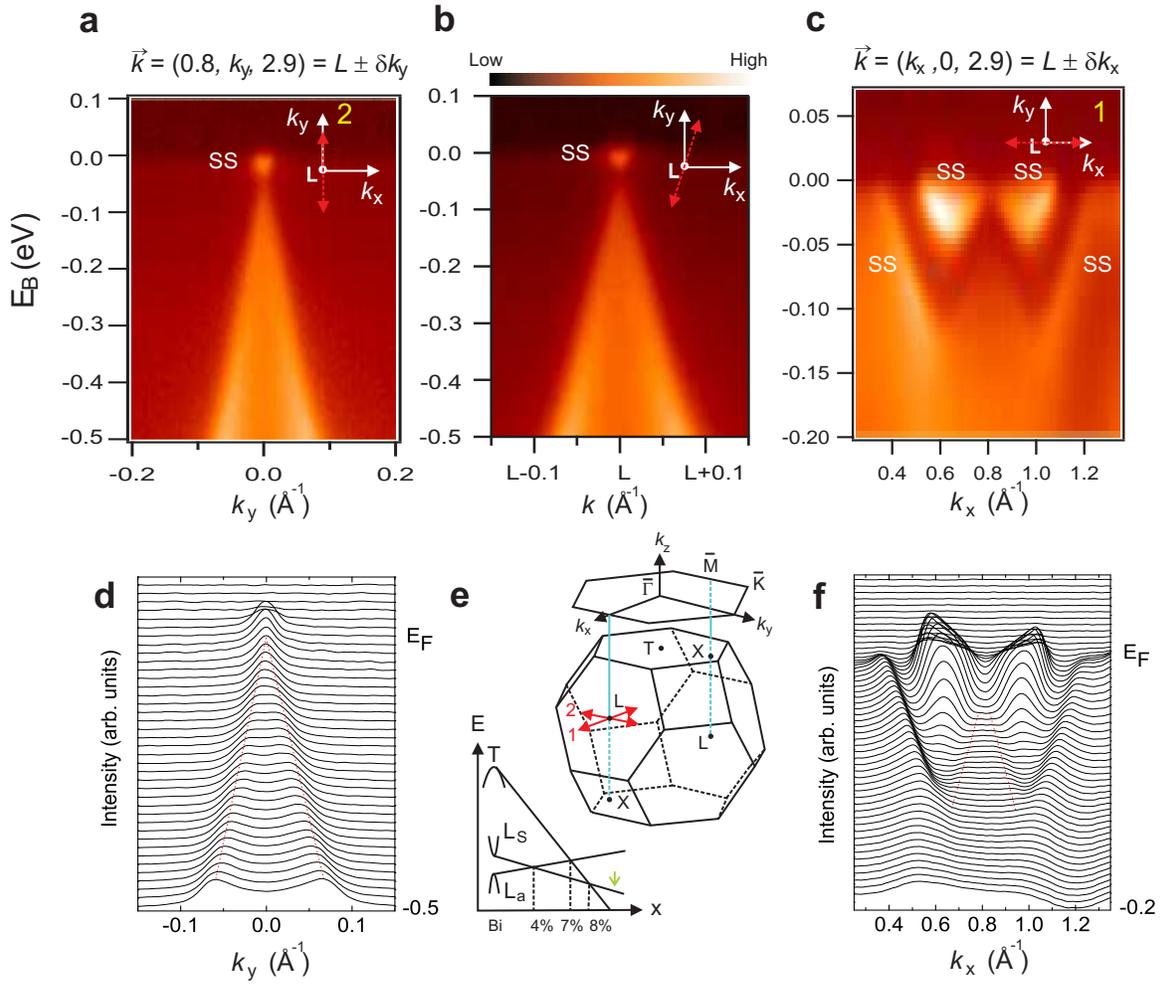}
\caption{\label{fig:BiSb_Fig2} Hsieh $et$ $al.$, NATURE 452, 970 (2008).}
\end{figure}

\newpage

\begin{figure}
\includegraphics[scale=0.65,clip=true, viewport=0.0in 0in 10.0in 7.7in]{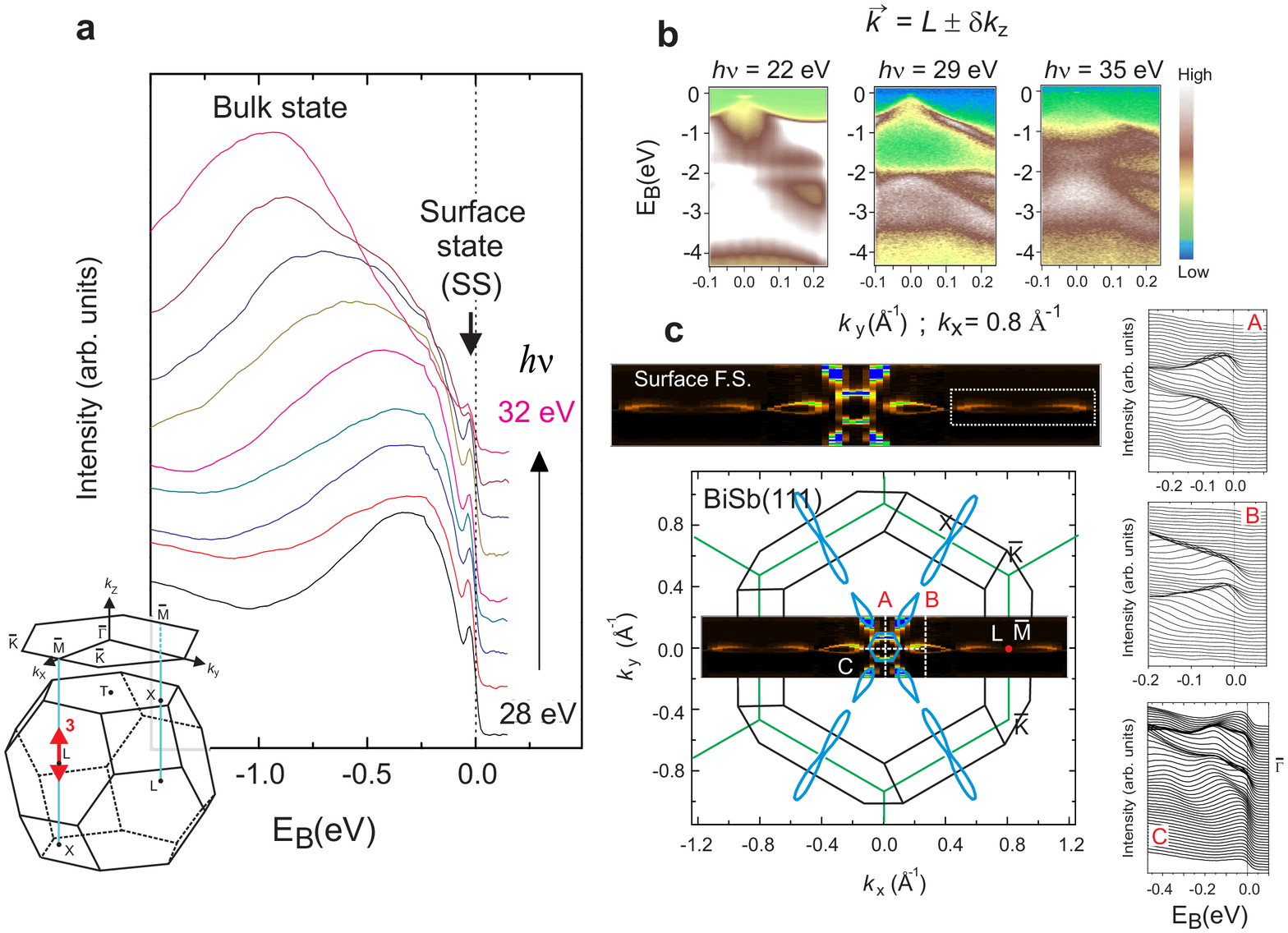}
\caption{\label{fig:BiSb_Fig2} Hsieh $et$ $al.$, NATURE 452, 970 (2008).}
\end{figure}

\newpage

\begin{figure}
\includegraphics[scale=0.65,clip=true, viewport=0.0in 0in 10.5in 7.7in]{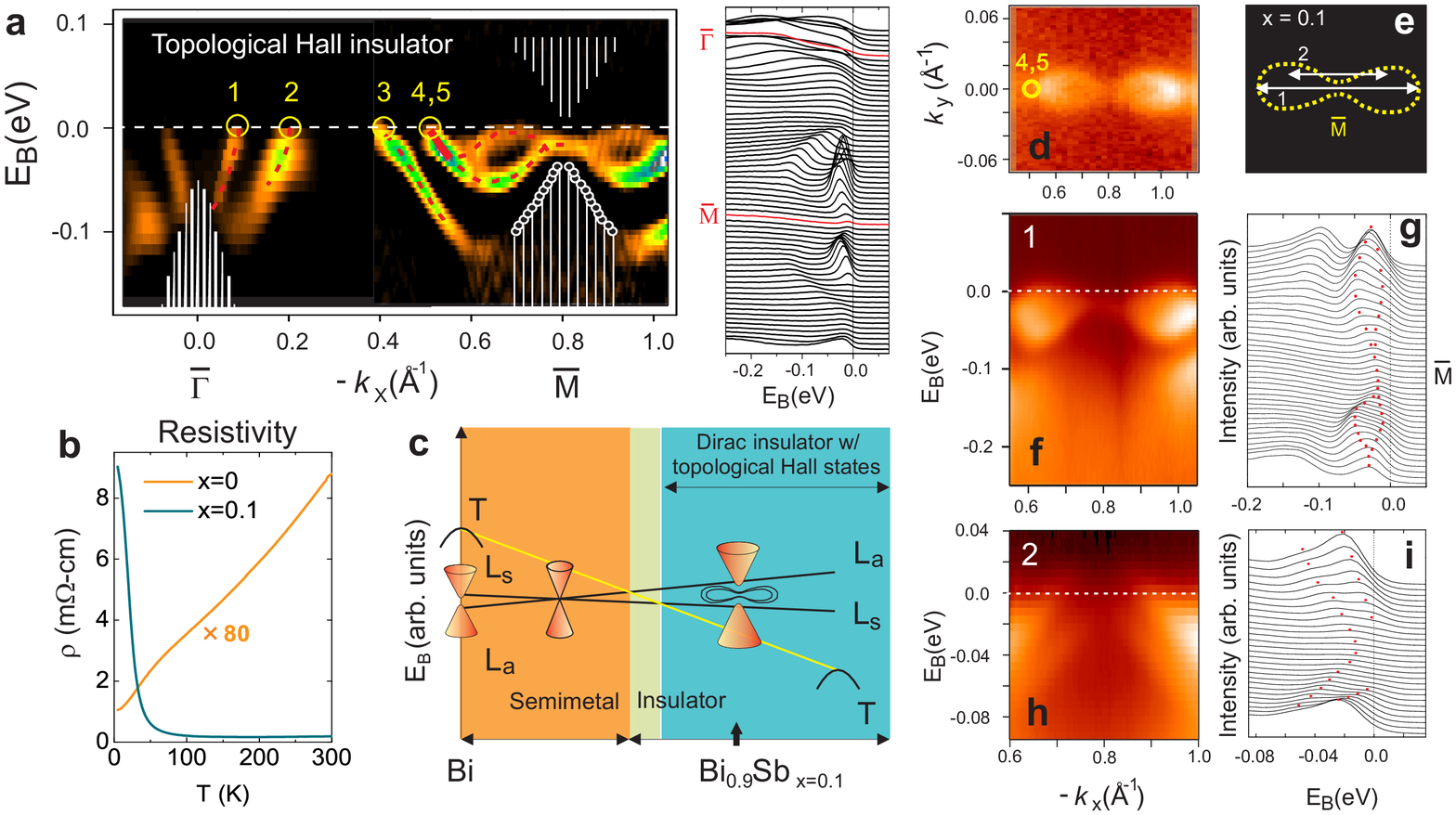}
\caption{\label{fig:BiSb_Fig2} Hsieh $et$ $al.$, NATURE 452, 970 (2008).}
\end{figure}

\end{document}